\documentclass[12pt,preprint]{aastex}

\slugcomment{Not to appear in Nonlearned J., 45.}

\shorttitle{Collapsed Cores in Globular Clusters}
\shortauthors{Djorgovski et al.}

\begin{document}


\title{Coronagraphic demonstration experiment using aluminum mirrors for space infrared astronomical observations}


\author{S. Oseki\altaffilmark{1}, S. Oyabu\altaffilmark{1}, D. Ishihara\altaffilmark{1}, K. Enya\altaffilmark{2}, K. Haze\altaffilmark{2}, T. Kotani\altaffilmark{3}, H. Kaneda\altaffilmark{1}, M. Nishiyama\altaffilmark{1}, L. Abe\altaffilmark{4}, T. Yamamuro\altaffilmark{5}}
%%\affil{}
\email{oseki@u.phys.nagoya-u.ac.jp}

%%\author{C. D. Biemesderfer\altaffilmark{4,5}}
%%\affil{National Optical Astronomy Observatories, Tucson, AZ 85719}

%%\and

%%\author{R. J. Hanisch\altaffilmark{5}}
%%\affil{Space Telescope Science Institute, Baltimore, MD 21218}


\altaffiltext{1}{Graduate School of Science, Nagoya University, Furo-cho, Chikusa-ku, Nagoya 464-8602, Japan}
\altaffiltext{2}{Institute of Space \& Astronautical Science, Japan Aerospace Exploration Agency, 3-1-1 Yoshinodai, Chuo-ku, Sagamihara, Kanagawa 252-5210, Japan}
\altaffiltext{3}{National Astronomical Observatory of Japan, 2-21-1 Osawa, Mitaka, Tokyo 181-8588, Japan}
\altaffiltext{4}{Laboratoire Universitaire d'Astrophysque de Nice, UMR 6525, Parc Valrose, F-06108 Nice, France}
\altaffiltext{5}{Optcraft, 3-26-8 Aihara, Sagamihara, Kanagawa 229-1101, Japan}


\begin{abstract}
For future space infrared astronomical coronagraphy, we perform experimental studies on the application of aluminum mirrors to a coronagraph. Cooled reflective optics is required for broad-band mid-infrared observations in space, while high-precision optics is required for coronagraphy. For the coronagraph instrument originally proposed for the next-generation infrared astronomical satellite project SPICA (SCI: SPICA Coronagraph Instrument), we fabricated and evaluated the optics consisting of high-precision aluminum off-axis mirrors {\bf with diamond-turned surfaces}, and conducted a coronagraphic demonstration experiment using the optics with a coronagraph mask. We first measured the wave front errors (WFEs) of the aluminum mirrors with a He-Ne Fizeau interferometer to confirm that the power spectral densities of the WFEs satisfy the SCI requirements. Then we integrated the mirrors into an optical system and evaluated the overall performance of the system. As a result, we estimate the total WFE of the optics to be 33 nm (rms), each mirror contributing {\bf 10$-$20 nm (rms) for the central 14 mm area of the optics}, and obtain a contrast of $10^{-5.4}$ as a coronagraph in the visible light. At a wavelength of 5 $\mu$m, the coronagraphic system is expected to achieve a contrast of $\sim10^{-7}$ based on our model calculation with the measured optical performance. Thus our experiment demonstrates that aluminum mirror optics is applicable to a highly WFE-sensitive instrument such as a coronagraph in space.

\end{abstract}





\section{Introduction}

It is important to evaluate the properties of extra-solar planets such as temperature, mass, age, and atmospheric composition by detailed observations to understand a planetary system formation history as well as a variety of planets. More than 1900 exoplanets have been discovered up to now (Extrasolar Planets Encyclopedia http://exoplanet.eu/catalog/). However, most of the exoplanets were found by using indirect methods such as Doppler and transit methods. Besides the fact that indirect methods are less effective for exoplanets located far from main stars, we still have limited information to characterize the properties of exoplanets using indirect methods. Thus, spatially resolved direct imaging is important to characterize various exoplanet systems. In particular mid-infrared (IR) spectral ranges are rich in exoplanet atmospheric gas lines and crucial for the characterizations of the systems. 

We need to consider a very high contrast in the flux between a main star and its planets when we measure the images of exoplanets directly. For example, when we observe Jupiter in the solar system from the outside, we have to overcome the contrast between the Sun and Jupiter which is $\sim10^{-9}$ in the visible light. On the other hand, the contrast in the mid-IR ($\sim10^{-6}$) is much lower than that in the visible light. Nevertheless we cannot observe such an object from the ground in the mid-IR due to the effect of Earth atmospheric absorption. Hence coronagraphic observations with space telescopes are indispensable for direct imaging of exoplanets in the mid-IR.

SCI (SPICA Coronagraph Instrument) is one of the focal-plane instruments originally proposed for SPICA (SPace Infrared telescope for Cosmology and Astrophysics), a next-generation infrared astronomical satellite project. We adopt a binary shaped pupil mask \citep[and its references]{2011AdSpR..48..323E}, because the mask has a high tolerance against the satellite pointing error and its optical performance does not depend on the observed wavelength \citep{2001astro.ph..1142S, van03a, van03b, van04, 2005ApOpt..44.1117K, 2005SPIE.5905..128K, 2007SPIE.6693E..0YB, 2007A&A...461..783E, 2008A&A...480..899E, 2009AdSpR..43..181H, 2010PASJ...62.1407E, 2011arXiv1108.4050C, 2011PASJ...63..873H, 2011SPIE.8146E..0QE, 2012arXiv1112.6301H, 2012SPIE.8442E..5CE}. It was confirmed from laboratory experiments that a contrast of $\sim10^{-7}$ in the visible light is obtained by using a lens optical system with the binary pupil mask \citep{2012arXiv1112.6301H}. In addition, they established a production method of free-standing masks, and developed various types of masks designed for partially obscured pupils.

We require a high-accuracy optical system to realize direct imaging of exoplanets, because coronagraph capabilities strongly depend on the wave-front error (WFE) of the optical system. For mid-IR observations in space, we have to cool the whole instrument to low temperatures in order to minimize the thermal radiation from a telescope and instruments. Considering the necessity of a broad-band coverage of mid-IR wavelengths as well as cooling of an optical system, mirror optics is more appropriate than lens optics. In order to maintain optical alignment when we cool the optical system, the material of mirrors should be the same as that of the instrument structure, which realizes uniform thermal contraction of the whole optical system. Hence we develop and evaluate an off-axis aluminum mirror optical system to satisfy the requirements as a coronagraph instrument with the binary pupil mask developed for SCI.

We fabricated the mirrors by ultra-precision machining, because aluminum is difficult to grind and polish with high surface quality due to its mechanical properties such as the low Young's modulus. We measured the surfaces of the aluminum mirrors and estimated the WFEs from the difference between the designed and the measured mirror surfaces to verify that the mirror optics satisfies the SCI requirements. We then integrated the mirrors into the optical system with the binary pupil mask to evaluate the performance as a coronagraph. In the optical measurement, we use a visible light (He-Ne) laser, the wavelength of which is about 10 times shorter than the targeted observational wavelengths for SCI and thus more sensitive to optical misalignment, if any, of the system. Finally we evaluate the capability of the coronagraph at a wavelength of 5 $\mu$m, the shortest wavelength to be covered with SCI, based on our model calculation with the measured optical performance.

\section{Measurements}
\subsection{Aluminum mirrors}

%% In a manner similar to \objectname authors can provide links to dataset
%% hosted at participating data centers via the \dataset{} command.  The
%% second curly bracket argument is printed in the text while the first
%% parentheses argument serves as the valid data set identifier.  Large
%% lists of data set are best provided in a table (see Table 3 for an example).
%% Valid data set identifiers should be obtained from the data center that
%% is currently hosting the data.
%%
%% Note that AASTeX interprets everything between the curly braces in the 
%% macro as regular text, so any special characters, e.g. "#" or "_," must be 
%% preceded by a backslash. Otherwise, you will get a LaTeX error when you 
%% compile your manuscript.  Special characters do not 
%% need to be escaped in the optional, square-bracket argument.

We fabricated and tested 8 off-axis parabolic aluminum mirrors with two kinds of mirror surfaces as well as 1 flat aluminum mirror, all of which were made by Crystal Optics. Table~\ref{tbl-1} summarizes the designed focal and off-axis lengths as well as the material details for the off-axis parabolic and flat aluminum mirrors. The mirrors in the top 3 rows have the same surface shapes but with different aluminum materials.

\begin{table}[!ht]
%\begin{center}
\caption{Designed optical parameters and the material of the aluminum mirrors}\label{tbl-1}
\begin{tabular}{cccc}
\tableline\tableline
ID & focal length [mm] & off-axis length [mm] & material \\
\tableline
A1,B1 & 250 & 76 & A5052O    \\
V2A1, V2A2 & 250 & 76 & A6061T     \\
V2R1, V2R2 & 250 & 76 & RSA6061T6   \\
A2, B2 & 500 & 152 & A5052O    \\
F1 & - (flat mirror) & - (flat mirror) & A5052O     \\
\tableline
\end{tabular}
%% Text for table notes should follow after the \enddata but before
%% the \end{deluxetable}. Make sure there is at least one \tablenotemark
%% in the table for each \tablenotetext.
\vskip 2pt
\footnotesize{The surface size of all the mirrors is 50 mm $\times$ 50 mm.}
%\end{center}
\end{table}

\begin{figure}[h]
\begin{center}
\includegraphics[width=8cm]{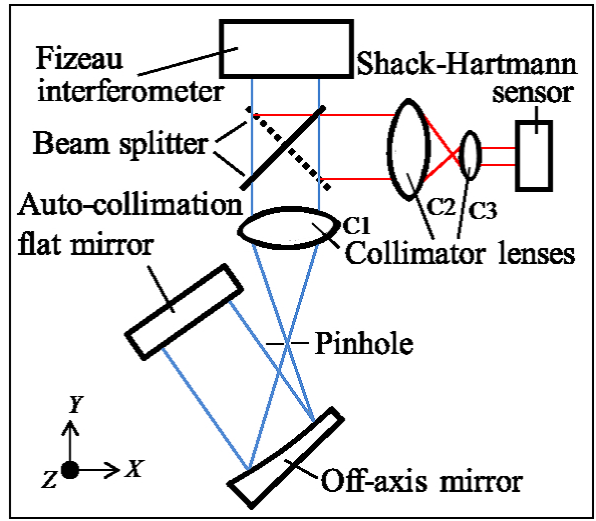}
\caption{Measurement system for off-axis parabolic mirrors. An off-axis mirror is mounted on a 5-axis optical adjustment stage. Additional optics with a Shack-Hartmann sensor is introduced for a coarse optical adjustment of the off-axis mirror.}
\label{fig:os01}
\end{center}
\end{figure}

We measure the surface figure errors of the 9 aluminum mirrors by a He-Ne (633 nm) Fizeau interferometer (FUJINON F601; Fizeau lenses with the clear aperture diameter 60 mm). Figure~\ref{fig:os01} shows the configuration of the measurement system for the parabolic mirrors, where additional optics is incorporated with the surface figure accuracy of $<\lambda$/20 (PV) for each optical component. We use the collimator lens C1 (effective diameter 50 mm, focal length 250 mm) so that the in-coming light focuses on the focal point of the off-axis parabolic mirror. The off-axis parabolic mirror is mounted on the 5-axis (X/Y/Z-shift and X/Z-tilt; see Figure 1) optical adjustment stage. We place an auto-collimation flat mirror in front of the off-axis mirror to reflect the light back to the interferometer. 

As a first step, we set a pinhole mask with a diameter of 1 mm at the designed focal position of the off-axis mirror. We adjusted the tilt angles of the off-axis mirror and the auto-collimation flat mirror, and alignment of the interferometer and C1. Then we performed shift adjustments of the off-axis mirror using the additional optics with a Shack-Hartmann sensor. This sensor consists of a micro lens array (lens pitch 300 $\mu$m, focal length 18.8 mm) and a CCD camera (570 $\times$ 570 pixels with the pixel scale 4.4 $\mu$m/pixel). The collimator lenses, C2 (effective diameter 50 mm, focal length 150 mm) and C3 (effective diameter 10 mm, focal length 20 mm), are used to adjust the beam diameter. With the Shack-Hartmann sensor, we coarsely adjusted the alignment of the off-axis mirror to the WFE of accuracy $\sim2.5$ $\mu$m (rms) which was estimated by the image distortions on the CCD camera. Then we performed a finer adjustment of the off-axis mirror alignment based on a fringe pattern with a Fizeau interferometer with the accuracy of the X tilt $<$1 arcmin, the Z tilt $<$20 arcsec, and the X, Y, and Z shifts $<$3 $\mu$m which are the resolutions of the optical adjustment stage.

We obtained the surface figure errors of the mirrors from fringe patterns with the Fizeau interferometer. The WFE (i.e., twice the surface figure error) maps of all the mirrors are shown in Figure~\ref{fig:os03}. For the off-axis mirrors in the top 3 rows in Table ~\ref{tbl-1}, we could not measure the whole square areas of 50 $\times$ 50 mm$^2$ but circular areas of 42 mm in diameter, because the aperture of the interferometer Fizeau lens is obscured by the fore optics, which limits the measurable areas. As for the other off-axis mirrors, we measured the whole square areas of 50 $\times$ 50 mm$^2$. For the flat mirror, we could not measure the edges of the square of 50 $\times$ 50 mm$^2$ with the 60-mm-diameter flat Fizeau lens. As explained later, only the central portion of each mirror is important for a coronagraph demonstration experiment, and therefore the coverage of the whole area is not necessary for that purpose. 

	It should be noted that the errors measured with the interferometer include not only the surface figure errors but also optical alignment errors; the latter might be significant for the off-axis mirror measurements. We therefore calculated the WFEs of the measurement system for the off-axis parabolic mirrors using the ray-tracing software ZEMAX and estimated the sensitivity of the alignment errors. To simplify the calculation we consider that alignment errors are caused by the resolution of the 5-axis optical adjustment stage. Thus there are 32 combinations of alignment errors of both plus and minus directions for each axis. We below choose the worst case as a possible contribution of the alignment error to the off-axis mirror measurement.

\subsection{Optics for a coronagraph demonstration experiment}
\begin{figure}[h]
\begin{center}
\includegraphics[width=8cm]{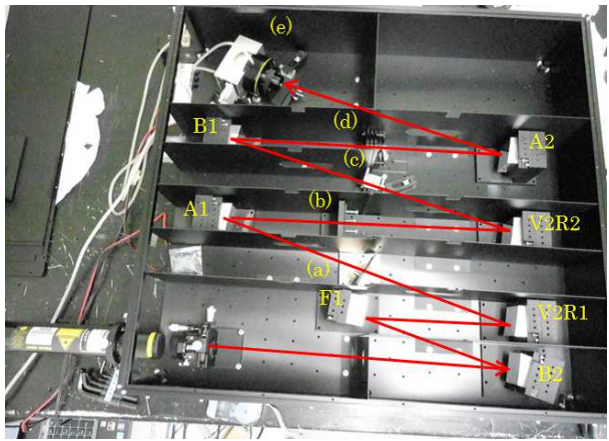}
\caption{Aluminum mirror optics for a coronagraph demonstration experiment. The He-Ne laser beam from the bottom left is focused through the optics onto the CCD camera placed at the top left.}
\label{fig:os02}
\end{center}
\end{figure}

Figure~\ref{fig:os02} shows the whole optics for the coronagraph demonstration experiment, configured on the optical bench (700 $\times$ 730 mm$^2$) for the SCI testbed, PINOCO \citep{2012SPIE.8442E..5CE}. We integrated six off-axis parabolic mirrors and a flat mirror with adjustable mirror holders into the optical system. We selected the mirrors, based on the result of the WFE measurement, i.e., mirrors with smaller WFEs regardless of the aluminum properties, on the surface because we did not observe any systematic difference in the surface figure error between the different materials. From the bottom on the left-hand side in Figure~\ref{fig:os02}, we input the He-Ne laser through a spatial filter. On the top left-hand side, we set a CCD camera (BJ-40L, BITRAN) of 772 $\times$ 580 pixels with the pixel size of 5.4 $\mu$m/pixel. 

	At position (a) in Figure~\ref{fig:os02}, the first focal point, a pinhole of 0.4 mm in diameter is placed to remove the scatter light from the fore optics. At position (b), the first pupil, either a coronagraph mask or a circular aperture mask is placed to control the point spread function (PSF). As a coronagraph mask, we use a free standing checker board mask made of copper with the following specifications: the contrast of 10$^{-10}$, the inner working angle (IWA) and the outer working angle (OWA) of 5.4 and 50 $\lambda/D$, respectively, the optical throughput of $\sim$0.34, and the aperture size of 10 $\times$ 10 mm$^2$ \citep{2012SPIE.8442E..5CE, 2012arXiv1112.6301H}. The diameter of the circular aperture is set to be 14 mm, which is the same as the diameter of the outer circular envelope of the coronagraph mask. At position (c), the second focal point, a focal plane mask is placed to reduce the central intensity of the PSF, which is constructed by two crossed blades with 0.1 mm thickness for each. At position (d), the second pupil, three ND filters of 10$^{-6}$ attenuation per filter are placed to reduce the total intensity of the beam, which are tilted by ~5 degrees with respect to the optical axis for stray-light rejection. Finally at position (e), the third focal point, the cooled CCD camera is placed on the focus adjustment stage.

We first measured the PSF image of the optics with the circular aperture. To derive the whole PSF, we took 20 short-exposure (0.01 s) frames and 20 long-exposure (30 s) frames. The former is to measure the intensity of the PSF center, while the latter is to measure the intensity of the PSF skirts. For a half of the latter measurement, we placed the focal plane mask at position (c) in Figure~\ref{fig:os02}. We then measured the PSF image of the optics with the coronagraph mask. Because the optical throughput of the coronagraph mask is lower than that of the circular aperture, we used longer exposure times, 0.04 s and 60 s, for the short-exposure and long-exposure frame measurements, respectively. Except for the difference in the exposure time, measurement procedures for the PSF with the coronagraph mask are the same as those for the PSF with the circular aperture. Hence we took six kinds of datasets, three with the circular aperture and three with the coronagraph mask, each three consisting of short exposure and long exposure without the focal plane mask and long exposure with the focal plane mask. Below we use the averaged image for each dataset and estimate intensity errors from the standard deviation.

\section{Result}
\subsection{Evaluation of the aluminum mirrors}
\begin{figure}[h!]
\begin{center}
\includegraphics[width=6cm]{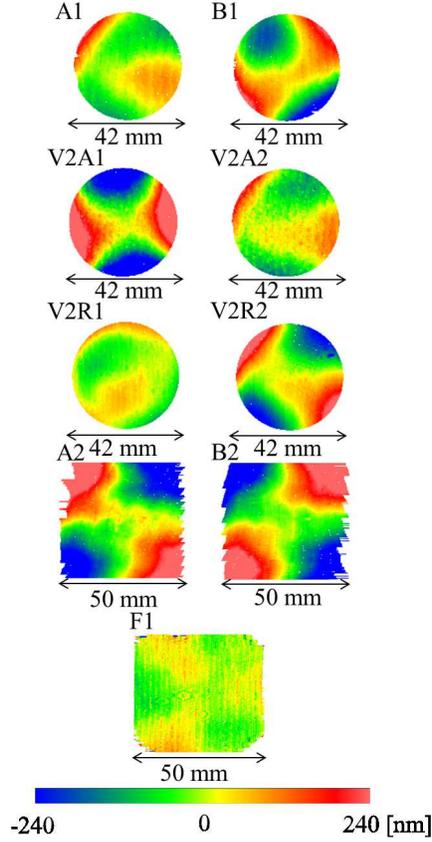}
\caption{WFE maps of all the aluminum mirrors. The color scales are common among the maps. The mirror surface of each mirror is a 50 $\times$ 50 mm square. For the top 6 off-axis mirrors and the bottom 1 flat mirror, outer boundaries are limited by the aperture of the measurement optics. }
\label{fig:os03}
\end{center}
\end{figure}

\begin{figure}[h!]
\begin{center}
\includegraphics[width=10cm]{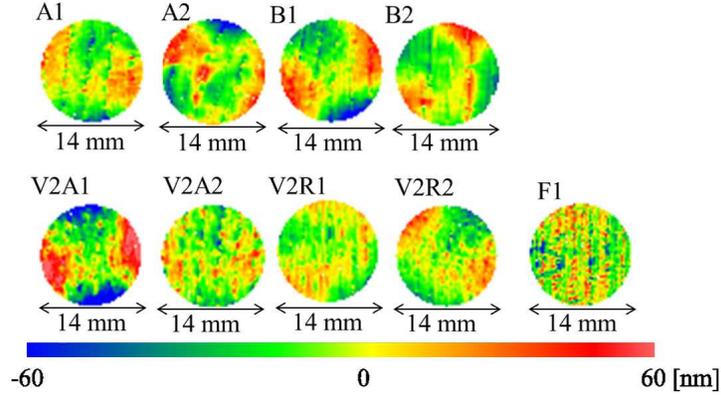}
\caption{WFE maps of the central circular area of 14 mm in diameter for the aluminum mirrors used in the coronagraph demonstration experiment.}
\label{fig:os04}
\end{center}
\end{figure}

Figure~\ref{fig:os03} shows the WFE maps of all the aluminum mirrors fabricated and measured in the present study. As mentioned above, for the top 6 off-axis mirrors and the flat mirror, outer circular boundaries are limited by the aperture of the measurement optics, while the whole areas of 50 $\times$ 50 mm$^2$ are measured for the other 2 off-axis mirrors although they show data losses due to poor surface quality at the edges. The overall WFE distributions of the off-axis parabolic mirrors are apparently dominated by astigmatism. Table~\ref{tbl-2} summarizes the WFE values of all the mirrors. The standard deviations are obtained by multiple WFE measurements. It should be noted that the WFEs include not only the surface figure errors themselves but also the optical alignment errors in the measurement system, especially those for the off-axis mirrors. Therefore possible alignment errors of the off-axis mirror measurement system was evaluated with the ray-tracing analysis as described above and shown in the table. Figure~\ref{fig:os04} shows the WFE maps of the central circular areas of 14 mm in diameter for the 9 aluminum mirrors. These partial areas actually contribute to the total WFE of the optical system for the coronagraph demonstration experiment. The WFE values of the surface areas in Figure~\ref{fig:os04} are also summarized in Table~\ref{tbl-2}, which are confirmed to be as small as about {\bf 10$-$20 nm (rms)}.

\begin{table}[!ht]
\caption{WFEs of all the mirrors\label{tbl-2}}
\begin{tabular}{cccc}
\tableline\tableline
ID & WFE for the full area measured [nm] (rms) & WFE for the partial area$^{\rm a}$[nm] (rms) \\
\tableline
A1 & 77 $\pm$ 5 (26) & 17 $\pm$ 1 (1)    \\
B1 & 131 $\pm$ 25 (26) & 19 $\pm$ 1 (1)    \\
V2A1 & 247 $\pm$ 10 (26) & 33 $\pm$ 2 (1) \\
V2A2 & 85 $\pm$ 2 (26) & 16 $\pm$ 2 (1) \\
V2R1 & 53 $\pm$ 4 (26) & 12 $\pm$ 1 (1)    \\
V2R2 & 148 $\pm$ 5 (26) & 17 $\pm$ 1 (1)    \\
A2 & 129 $\pm$ 5 (36) & 20 $\pm$ 2 (2)    \\
B2 & 125 $\pm$ 13 (36) & 17 $\pm$ 1 (2)    \\
F1 & 40 $\pm$ 2 & 21 $\pm$ 1 \\
\tableline
\end{tabular}
%% Text for table notes should follow after the \enddata but before
%% the \end{deluxetable}. Make sure there is at least one \tablenotemark
%% in the table for each \tablenotetext.
%\tablecomments{The numbers in the parentheses are possible contributions of the alignment errors of the off-axis mirror measurement system, which is evaluated using the ray-tracing software ZEMAX.}
\vskip 2pt
\footnotesize{The numbers in the parentheses are possible contributions of the alignment errors of the off-axis mirror measurement system, which is 
evaluated using the ray-tracing software ZEMAX.}

\footnotesize{$^{\rm a}$The central circular area of diameter 14 mm, which is used in the coronagraph demonstration. }
\end{table}

\begin{figure}[h]
\begin{center}
\includegraphics[width=11cm]{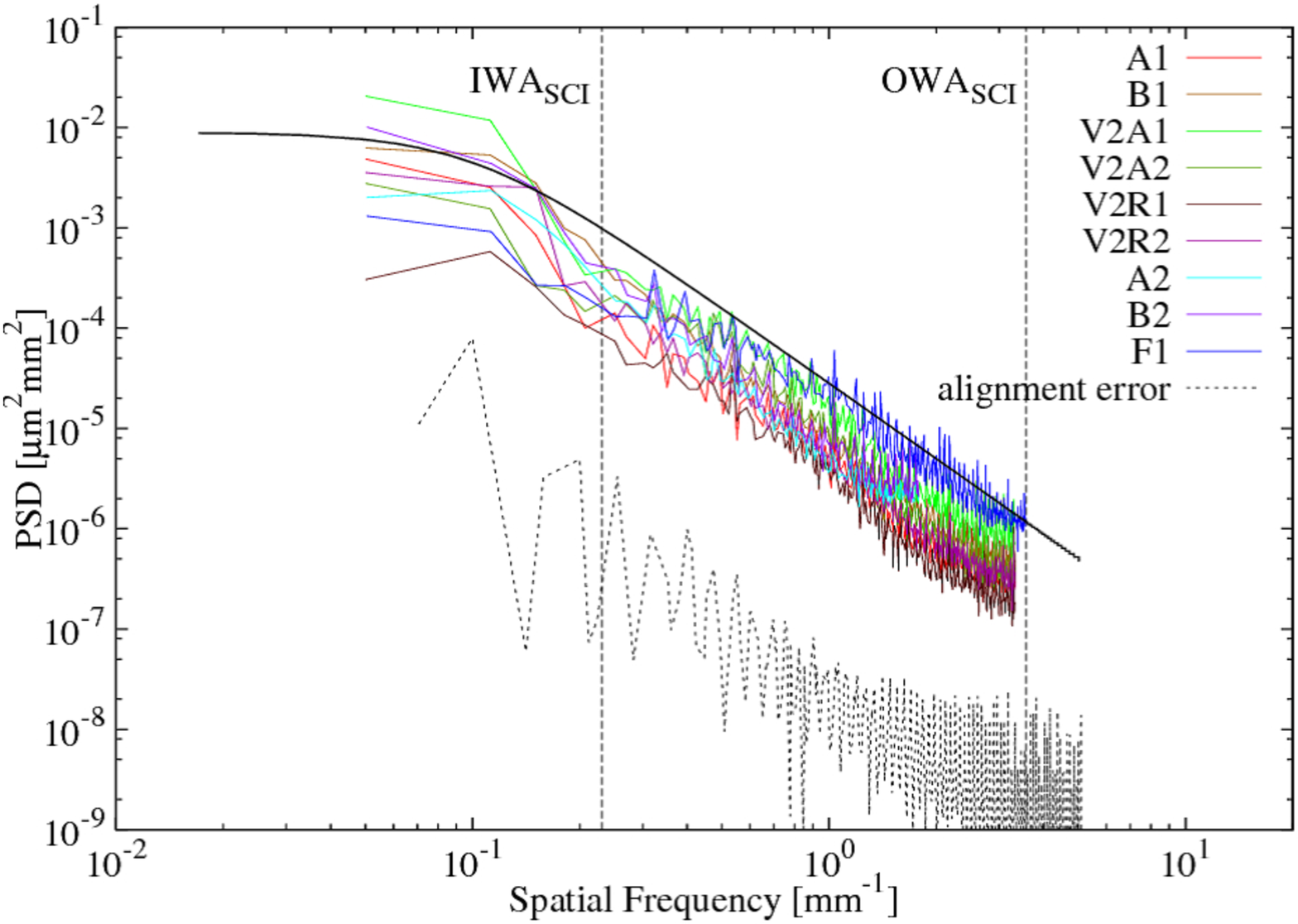}
\caption{PSDs calculated from the WFE maps in Figure~\ref{fig:os04} for the mirrors used in the coronagraph demonstration experiment. The black solid curve indicates the SCI requirement on the WFE PSD of each mirror.}
\label{fig:os05}
\end{center}
\end{figure}

Figure~\ref{fig:os05} shows the PSDs calculated from the WFE maps in Figure~\ref{fig:os04} for the mirrors used in the coronagraph demonstration experiment. The minimum and maximum limits of the spatial frequency range are determined by the diameter of the measured surface and the spatial resolution of the Fizeau interferometer, respectively. Black dotted curve in the bottom indicates the expected contribution of the alignment error in the off-axis mirror measurement to the measured WFE, which is estimated by the ray-tracing simulation. The black solid curve is derived from the SCI requirement on the WFE of each optical component. As a model PSD, we adopt the following function \citep{2006ApJ...638..488B}:
\[PSD(\rho) = A/(1+\rho/\rho_c)^x,\]
where $\rho$ is the spatial frequency and $A$, $\rho_c$, and $x$ are constant parameters. We determined $\rho_c$ and $x$ by fitting the above function to the PSD averaged over the mirrors ($\rho_c$ = 0.1 mm$^{-1}$, $x$ = 2.5), while parameter $A$ is determined for the PSD to give {\bf 30 nm (rms)}, the SCI requirement on the WFE of each mirror, which is based on the SPICA science requirement for detection of Jupiter-type planets with a coronagraph in the mid-IR. Two vertical broken lines indicate the IWA and the OWA of SCI. As can be seen in the figure, the higher spatial frequency end is not sufficiently covered by the present WFE measurement to judge whether or not the aluminum mirror optics meets the SCI requirement for all the frequency range. We verify this later in the coronagraph demonstration experiment. Figure~\ref{fig:os05} indicates that all the PSDs mostly satisfy the SCI requirement, although the PSDs in two regions (F1: 0.9 -- 3.3 mm$^{-1}$, B1, V2A1, B2: 0.06 -- 0.15 mm$^{-1}$) slightly exceed the SCI requirement. 

\subsection{Coronagraph demonstration with the aluminum mirror optics}
\begin{figure}[h]
\begin{center}
\includegraphics[width=6cm]{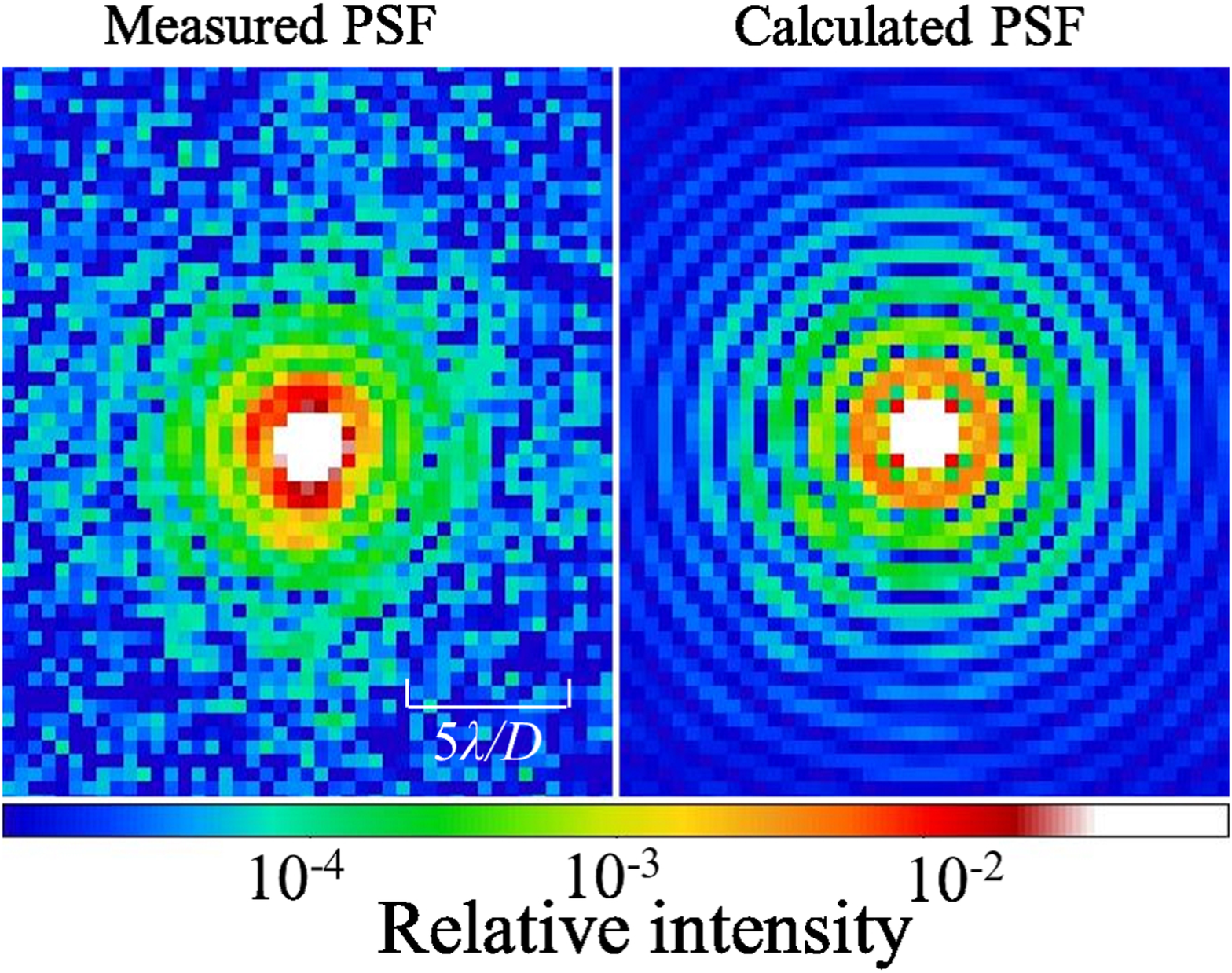}
\caption{(Left) Measured and (Right) calculated PSFs of the optical system for the coronagraph demonstration experiment. {\bf The former PSF is taken by the short-exposure measurements.} In both images, the intensity is normalized by the peak.}
\label{fig:os06}
\end{center}
\end{figure}

We integrated the 7 aluminum mirrors measured in section 3.1 into the optical system for the coronagraph demonstration experiment. In order to evaluate the total WFE of the whole optics, we measure the Strehl ratio of the PSF. Figure~\ref{fig:os06} shows the measured and calculated PSFs without the coronagraph mask (but with the circular mask). The peak intensity was obtained by fitting the Bessel function to the PSF. Then the Strehl ratio, $S$, i.e., the ratio of the peak intensity of the measured PSF to that of the calculated PSF, is estimated to be 0.89, which is related to the WFE $\sigma_{\rm {rms}}$ approximately by the following equation: 
\[S \sim 1-(2\pi/\lambda)^2\sigma_{\rm {rms}}^2\]
Since we used the He-Ne laser of a wavelength $\lambda$= 633 nm, $\sigma_{\rm {rms}}$ = 33 nm (rms). The total WFE of 33 nm (rms) meets the SCI requirement on the total WFE of 128 nm (rms) with a fair margin.

\begin{figure}[h]
\begin{center}
\includegraphics[width=6cm]{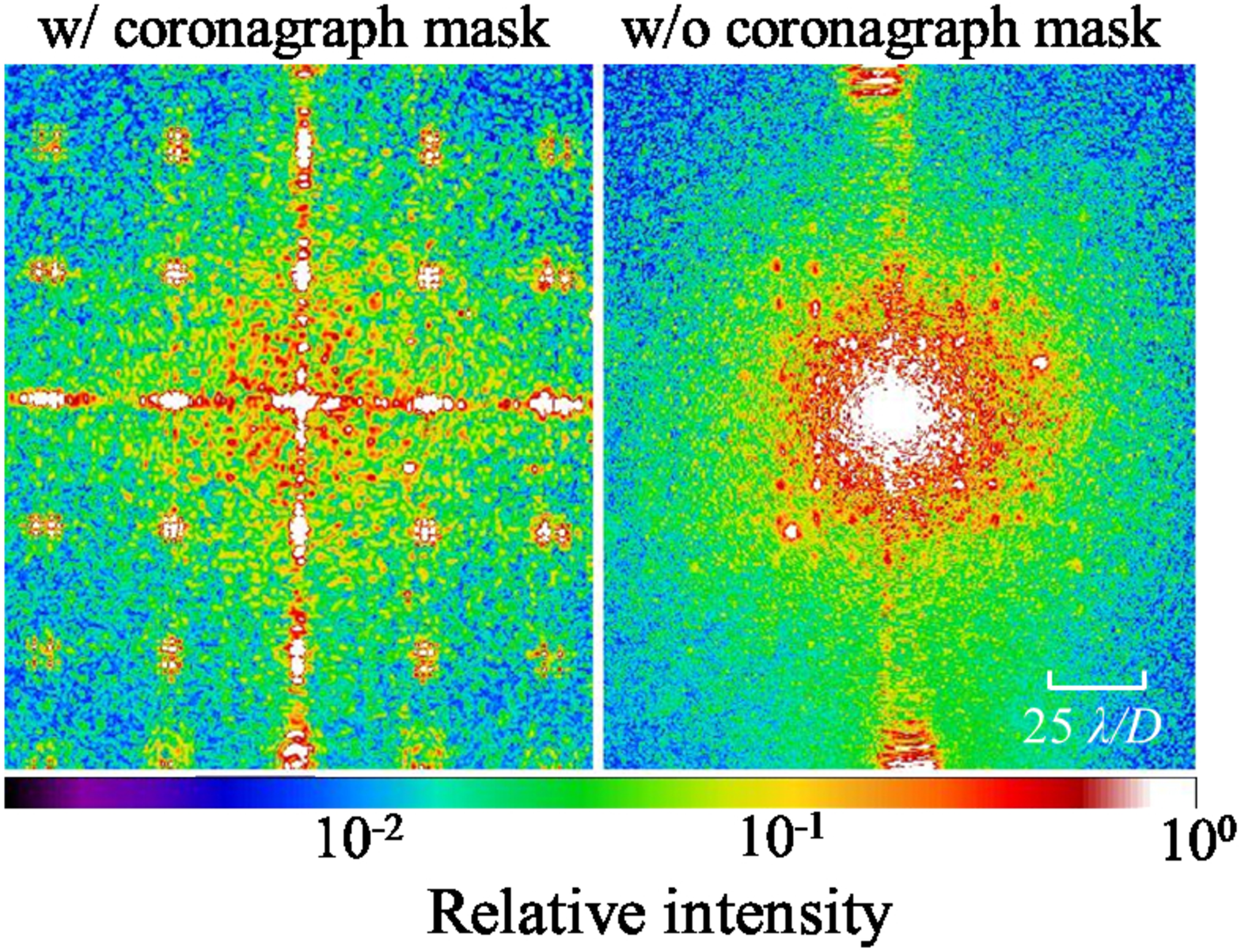}
\caption{PSFs measured in the coronagraph demonstration experiment with and without the coronagraph mask. {\bf The PSFs are taken by the long-exposure measurements}}
\label{fig:os07}
\end{center}
\end{figure}

Figure~\ref{fig:os07} shows the PSFs measured in the coronagraph demonstration experiment with and without a coronagraph mask. There are bright spots at the top and the bottom in the PSF without a coronagraph mask, which are located $\sim75$ $\lambda/D$ away from the center. This might be attributed to cutting traces on the aluminum mirror surfaces.  The above angular distance corresponds to the periodical structures of 186 $\mu$m in the length on the surfaces. We inspected the mirror surfaces by a microscope and found that these may be typical sizes of the aluminum cutting traces although the periodicity is not very clear. These bright spots are, however, located far enough from the center so that they do not affect the dark region (5.4 -- 50 $\lambda/D$: hereafter DR) created by the coronagraph mask.

\begin{figure}[h!]
\begin{center}
\includegraphics[width=8cm]{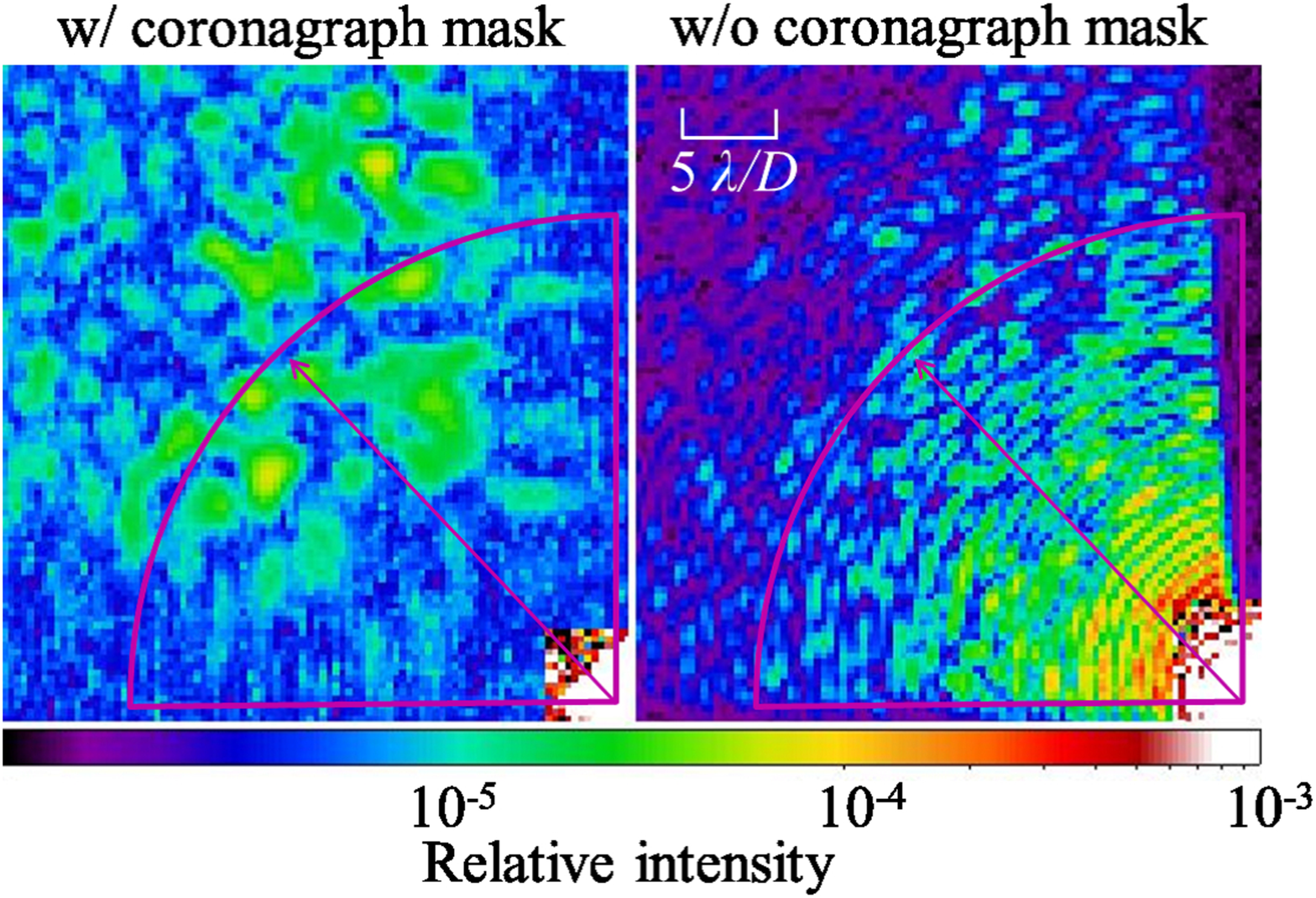}
\caption{Close-up images of the PSFs in Figure~\ref{fig:os07} but measured with the focal plane mask. The circular arc corresponds to the radius of 25 $\lambda/D$ from the center of the PSF.}
\label{fig:os08}
\end{center}
\end{figure}

Figure~\ref{fig:os08} shows the close-up images of the PSFs {\bf with and without the coronagraph mask} in Figure~\ref{fig:os07} but measured with the focal plane mask. To evaluate the contrast of the PSFs quantitatively, we create the radial profiles of the PSFs averaged azimuthally in the quadrant as shown in Figure~\ref{fig:os08}. The resultant PSF profiles are plotted in Figure 9, where the PSF with the coronagraph mask exhibits the contrast of $\sim10^{-5.4}$ for an angular distance range of 5.4 -- 25 $\lambda/D$. Around the IWA of SCI, the aluminum mirror optics with the coronagraph mask produces a contrast by 2 orders of magnitude better than that without the coronagraph mask. The PSF contrast measured between the IWA and OWA satisfies the SCI requirements, which implies that the PSDs of the WFEs at the high spatial frequency end, which were not evaluated by the interferometric measurement, might also meet the WFE requirements of SCI. 

\begin{figure}[h!]
\begin{center}
\includegraphics[width=9cm]{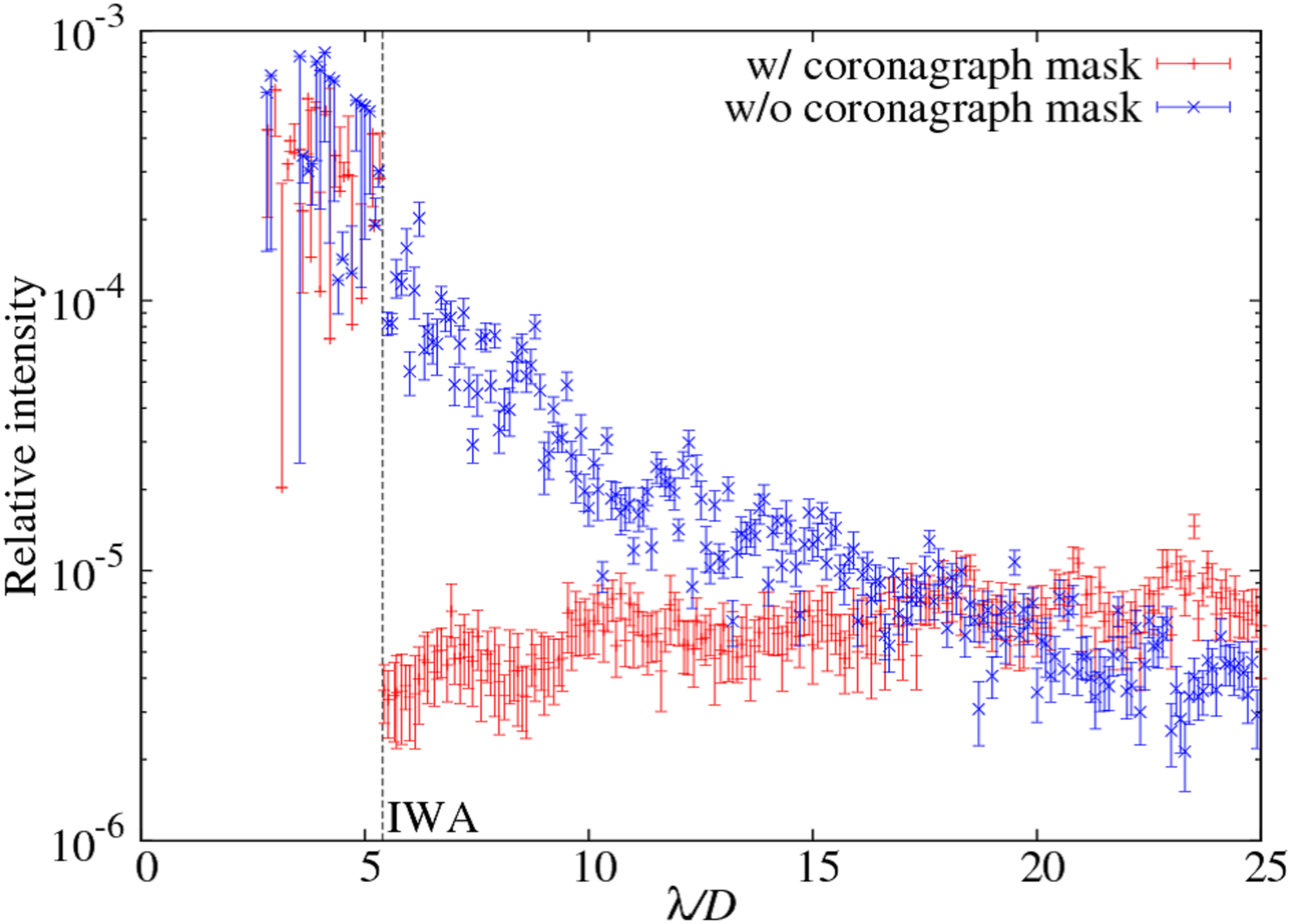}
\caption{Radial profiles of the PSFs measured with and without the coronagraph mask, in which the intensities of both profiles are normalized to unity at the center (i.e., $\lambda/D$ = 0).}
\label{fig:os09}
\end{center}
\end{figure}

\section{Discussion}
\subsection{Optical performance of the aluminum mirror optics}
As mentioned in section 2.1, the measured WFE maps include not only the surface figure errors of the mirror but also optical alignment errors in the off-axis mirror measurement. With the ray-trace analysis, we estimated the error map due to optical misalignment in the worst case conceivable within the positional uncertainties of the measurement configuration. Figure~\ref{fig:os05} shows the PSD of the error map thus estimated in the black dotted curve. As can be seen in the figure, the alignment error does not affect the measured WFE very much, but with about 10 percent levels at every spatial frequency.

	We estimated the total WFE of the aluminum mirror optics used in the coronagraph demonstration experiment from the measured Strehl ratio of 33 nm (rms) in section 3.1. On the other hand, the square root of the sum of the squared WFE of the individual mirrors amounts to the total WFE of 47 nm (rms). Therefore both estimates on the total WFE give results consistent to each other, considering the uncertainties of the optical alignment in the aluminum mirror optics. 

\subsection{Capability of the coronagraph with the aluminum mirror optics}
\begin{figure}[h!]
\begin{center}
\includegraphics[width=9cm]{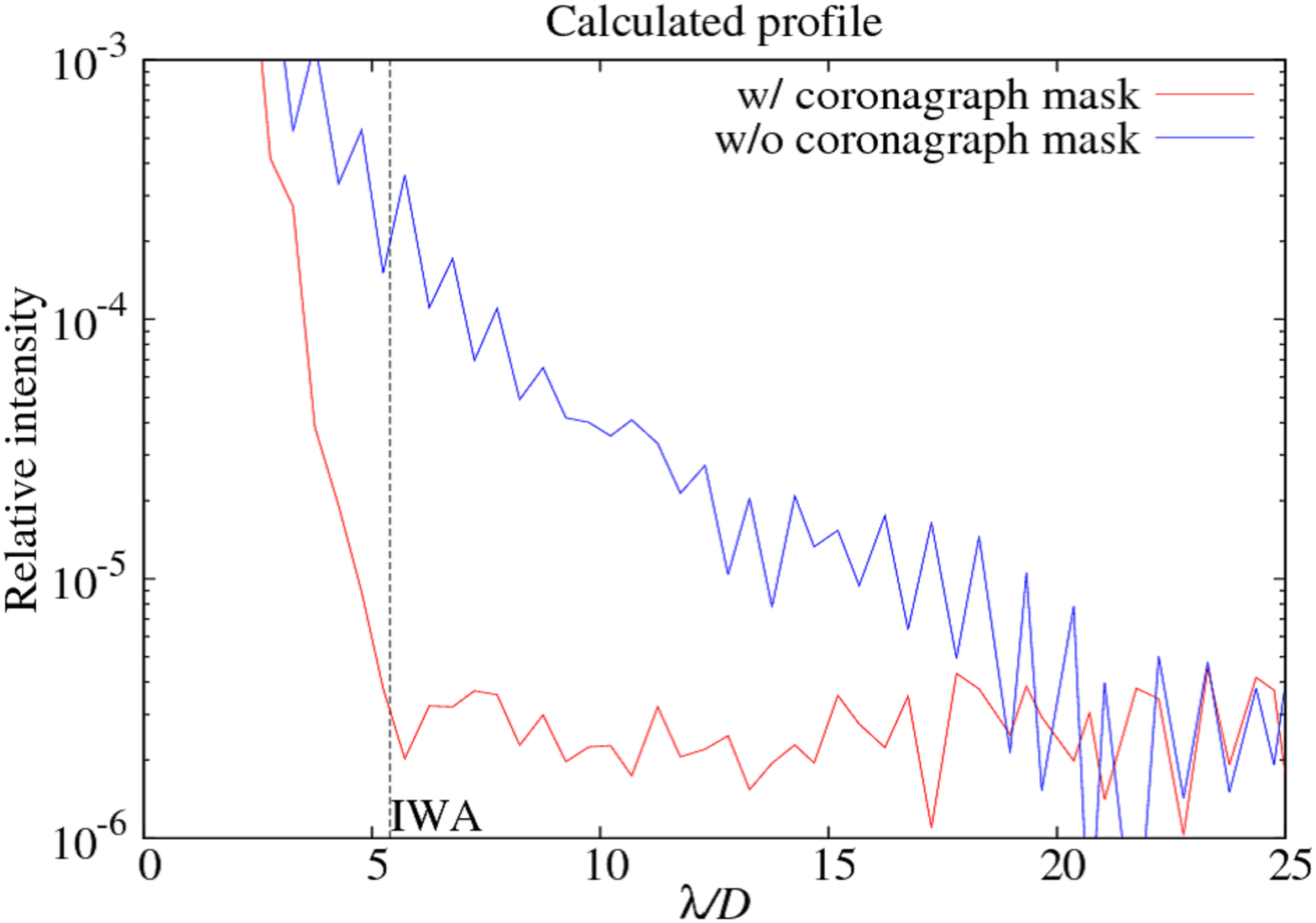}
\caption{Same as Figure~\ref{fig:os09}, but the profiles calculated with and without the coronagraph mask on the basis of the measured WFEs of the aluminum mirror optics. }
\label{fig:os10}
\end{center}
\end{figure}

Figure~\ref{fig:os10} shows the result of model calculation of the PSF profiles with and without the coronagraph mask at the He-Ne wavelength (632.8 nm). In the calculation, we consider the measured total WFE of the aluminum mirror optics, 33 nm (rms), assuming the PSD to be the same except the amplitude with that of the SCI requirement as shown in Figure~\ref{fig:os05}. Comparing the profiles in Figure~\ref{fig:os10} with those in Figure~\ref{fig:os09}, they show overall consistency with each other in the profile with the coronagraph mask; for example, contrast improvement by a factor of $\sim$100 around the IWA and the relative intensities of $\sim10^{-5.5}$ between the IWA and OWA with the coronagraph mask. 

\begin{figure}[h!]
\begin{center}
\includegraphics[width=9cm]{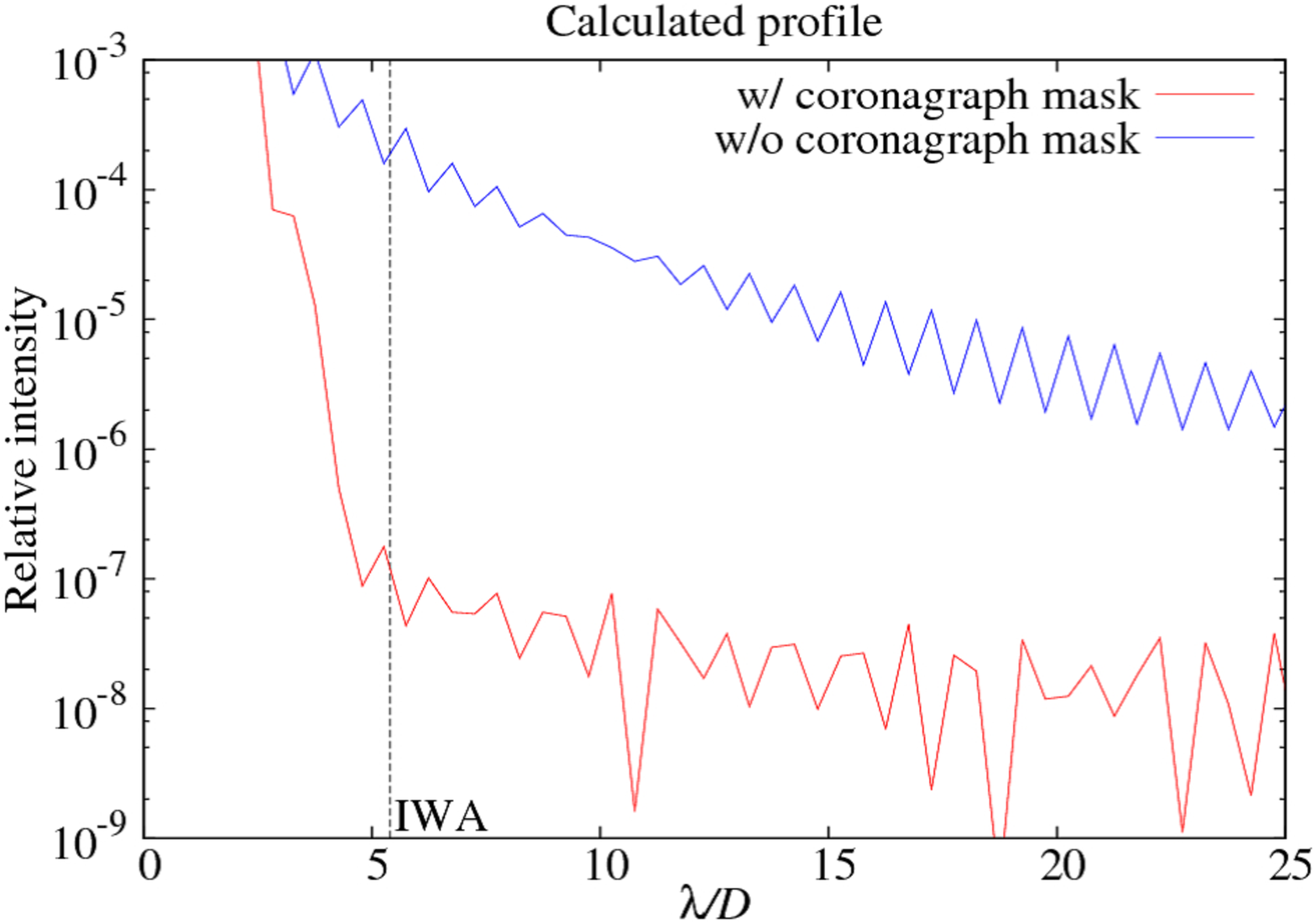}
\caption{Same as Figure~\ref{fig:os10}, but the profiles calculated for a wavelength of 5 $\mu$m instead of the He-Ne wavelength.}
\label{fig:os11}
\end{center}
\end{figure}

Using the same model calculation with the same aluminum off-axis mirror optics, we also show the profiles for an infrared wavelength of 5 $\mu$m in Figure~\ref{fig:os11}. As can be seen in the figure, the calculation predicts the contrast of $\sim10^{-7}$ around the IWA of SCI, and $10^{-7}$ -- $10^{-8}$ between the IWA and OWA of SCI. Since the surface figure errors of mirrors become relatively less important as the wavelength increases, we can achieve an even higher contrast at mid-IR wavelengths longer than the wavelength of 5 $\mu$m. Since the contrast between the Sun and Jupiter is $\sim10^{-6}$ in the mid-IR, we expect a direct mid-IR imaging of a Jupiter in an extra-solar system using the SCI which adopts the aluminum off-axis mirror optics with the coronagraph mask, as demonstrated in the present study.  

\section{Summary}
	We have studied the applicability of aluminum mirror optics for mid-IR astronomical coronagraphy in space, considering the platform for the SPICA coronagraph instrument, SCI. Cooled reflective optics is indispensable for high-sensitivity, broad-band mid-IR astronomical observations in space, while high-precision optics is required for a coronagraph instrument. To cool the whole instrument structure including mirror optics is crucial to suppress thermal background noise in space. As a cooled mirror material, aluminum is appropriate, because the aluminum is usually used for the instrument structure so that we can mitigate optical misalignment due to thermal distortion in cooling the instrument. However it is generally known that aluminum mirrors have difficulty in achieving high surface quality as compared to glass mirrors. 

	We fabricated and tested 8 off-axis aluminum mirrors as well as 1 flat aluminum mirror for a coronagraph demonstration experiment. We measured the WFE map of each aluminum mirror using a He-Ne Fizeau interferometer, and calculated the PSD. As a result, we confirm that all the mirrors satisfy the SCI requirements, {\bf although we are yet to cool the mirrors in order to verify that they do not show serious cryogenic distortions}. We then integrated 6 off-axis mirrors and 1 flat mirror into an optical system. We estimated the total WFE of the system to be 33 nm (rms) from the peak intensity ratio of the measured PSF to the calculated PSF (i.e., Strehl ratio). The WFE thus estimated shows a fair agreement with the total WFE calculated by the square root of the sum of the square of the WFE measured for each mirror. 

	Next, combining the binary-pupil coronagraph mask with the aluminum mirror optical system, we evaluated the optical performance as a coronagraph. From the PSF profile measured with the coronagraph mask, we obtained the contrast of $10^{-5.4}$ around the IWA of SCI and verified that the PSF contrast between the IWA and OWA satisfies the SCI requirements. We confirm that the coronagraph model calculation with the measured total WFE reproduces the measured PSF profile consistently. Finally the model calculation for an infrared wavelength of 5 $\mu$m shows the contrast of $\sim10^{-7}$ around the IWA of SCI, and $10^{-7}$ -- $10^{-8}$ between the IWA and OWA of SCI. Hence our study demonstrates that aluminum off-axis mirror optics is applicable to a highly WFE-sensitive instrument such as a coronagraph in space.

\appendix

\end{document}